\documentclass[fleqn,10pt]{wlscirep}

\usepackage{booktabs}
\usepackage{graphicx}
\usepackage{multirow}
\usepackage{amsmath,amssymb,amsfonts}
\usepackage{amsthm}
\usepackage{mathrsfs}
\usepackage[title]{appendix}
\usepackage{xcolor}
\usepackage{textcomp}
\usepackage{manyfoot}
\usepackage{algorithm}
\usepackage{algorithmicx}
\usepackage{algpseudocode}
\usepackage{listings}
\usepackage{subcaption}
\usepackage{tabularx}
\usepackage[utf8]{inputenc}
\usepackage[T1]{fontenc}

\theoremstyle{definition}

\title{Citation Structural Diversity: A Novel and Concise Metric Combining Structure and Semantics for Literature Evaluation}

\author[1]{Mingyue Kong}
\author[1,*]{Yinglong Zhang}
\author[2]{Likun Sheng}
\author[1]{Kaifeng Hong}

\affil[1]{College of Physics and Information Engineering, Minnan Normal University, Zhangzhou, Fujian 363000, China}
\affil[2]{Library, Minnan Normal University, Zhangzhou, Fujian 363000, China}

\affil[*]{Corresponding Author: Yinglong Zhang, E-mail: \href{mailto:zhang_yinglong@126.com}{zhang\_yinglong@126.com}}

\begin{abstract}
	As academic research becomes increasingly diverse, traditional literature evaluation methods face significant limitations,particularly in capturing the complexity of academic dissemination and the multidimensional impacts of literature. To address these challenges, this paper introduces a novel literature evaluation model of citation structural diversity, with a focus on assessing its feasibility as an evaluation metric. By refining citation network and incorporating both ciation structural features and semantic information, the study examines the influence of the proposed model of citation structural diversity on citation volume and long-term academic impact. The findings reveal that literature with higher citation structural diversity demonstrates notable advantages in both citation frequency and sustained academic influence. Through data grouping and a decade-long citation trend analysis, the potential application of this model in literature evaluation is further validated. This research offers a fresh perspective on optimizing literature evaluation methods and emphasizes the distinct advantages of citation structural diversity in measuring interdisciplinarity.
	
	\vspace{1em}
	
\end{abstract}

\begin{document}
\flushbottom
\maketitle
\section*{Introduction}
As global research activities continue to grow, the depth and breadth of academic inquiry expand, leading to an exponential increase in the volume of academic literature. In this context, the scientific and accurate assessment of the academic value and impact of literature has become a key issue in academia, research management, and policy-making. Literature evaluation metrics play a crucial role not only in measuring the impact of academic achievements \cite{ref1} but also in providing a foundation for the rational allocation of research resources \cite{ref2}, optimizing academic decision-making, and shaping research policies \cite{ref3}.

Traditional literature evaluation methods, such as citation counts \cite{ref4} and journal impact factors \cite{ref5}, are widely used in academic assessments due to their simplicity and intuitive results. However, the limitations of these indicators are becoming increasingly apparent. Firstly, traditional metrics typically focus solely on the direct academic impact of literature, failing to effectively reflect the complex academic relationships and knowledge dissemination patterns between literatures. Secondly, these indicators are weak in capturing disciplinary diversity and academic innovation, which may lead scholars to focus excessively on citation counts while overlooking the long-term value and originality of their academic contributions.

To overcome these limitations, citation network-based literature evaluation methods have emerged in recent years, establishing a new research paradigm \cite{ref6}. Citation networks treat academic papers as nodes and citation relationships as edges, creating a complex network that reflects academic dissemination and knowledge exchange. By structurally analyzing citation networks, researchers can uncover insights that traditional metrics often fail to identify. However, existing studies predominantly focus on the direct citation relationships between papers, neglecting the more complex structural features within citation networks. Structural diversity \cite{ref7}, an important indicator for measuring node influence in social networks \cite{ref8, ref9}, holds significant potential for application in citation network analysis. 

This endeavor aims to reveal the distinctive roles that structural diversity plays in knowledge dissemination, thereby offering a fresh perspective for evaluating literature. On the other hand, identifying highly innovative and interdisciplinary papers can be highly inspiring for individual researchers, as they are more likely to broaden their research horizons and perspectives, thereby better assisting them in enhancing their innovative capabilities. Furthermore, the semantic information contained in literature, as a carrier of knowledge, is crucial for evaluation. However, current evaluation models have not fully integrated semantic information, limiting the potential for multidimensional analysis. The study of combining semantic information with citation structural features to construct novel citation structural diversity models, thus enhancing the accuracy and comprehensiveness of literature evaluation, remains an unexplored area of research.

To address the aforementioned issues, we propose a novel and concise literature evaluation model: Citation Structural Diversity. By incorporating semantic information and citation structural features, our model strengthens the ability to represent complex relationships between literatures. Additionally, it integrates multidimensional information fusion, providing a more accurate and comprehensive approach to literature evaluation. The primary objective of this study is to validate the potential application of the proposed model in literature evaluation and its significant role in identifying interdisciplinarity. It is important to note that our model complements, rather than replaces, existing bibliometric indicators.

The main contributions of this paper are as follows:
\begin{enumerate}
	\item A literature evaluation model centered on structural diversity is proposed, offering a new perspective on revealing the multidimensional value of literature within academic dissemination networks. The proposed model is based on the following intuition: citing multiple sources from different disciplines reflects a researcher's integration of multidisciplinary knowledge, aiding in the formation of an interdisciplinary perspective and innovative outcomes.
	\item An innovative integration of the citation structural features and the semantic content of literature is introduced, creating a multidimensional, comprehensive evaluation framework that significantly enhances the ability to represent complex relationships between literatures. Our model maintains strong evaluation performance even in scenarios with limited data or incomplete networks.
	\item Extensive experiments have been conducted on real datasets. The experimental results validate the effectiveness of our model.
\end{enumerate}
\section*{Related Work}
\subsection*{Structural Diversity}
The theory of structural diversity originates from a deep understanding of the roles that strong and weak ties play within social networks. Granovetter \cite{ref10} introduced the "Strength of Weak Ties" theory, which demonstrates that weak ties are essential for connecting different communities and facilitating the flow of information across boundaries. Building on this, Lazega et al. \cite{ref11} expanded the concept with the "Structural Hole" theory. This theory argues that nodes in structural hole positions can access diverse information resources by bridging otherwise independent groups, thereby acting as key intermediaries in the flow of information. In 2012, Ugander et al. \cite{ref7} introduced the concept of "structural diversity" based on large-scale social network data. Their research, based on data from the Facebook social network, demonstrated that the number of relatively independent groups within an individual local network, or the structural diversity of the neighborhood, is strongly correlated with the breadth and depth of influence the individual experiences in major decision-making processes. In other words, the greater the number of connected components in a neighborhood network, the more opportunities an individual has to encounter heterogeneous information, making their behaviors or viewpoints more susceptible to diverse information sources. 

In practice, structural diversity has proven to be highly valuable across various fields. In social network analysis, it highlights the multiple roles of nodes and their critical function as bridges between communities \cite{ref9}. In information dissemination models, nodes with high structural diversity often link multiple paths of information flow, becoming central to diffusion \cite{ref12}. Moreover, in complex network evaluation, structural diversity describes the distribution of node functions, aiding in the understanding of the network's overall topological characteristics \cite{ref13}.

In terms of technical methods, the further development of this theory has been driven by improvements in three dimensions: undirected graphs, directed graphs, and parameter optimization. In undirected graph analysis, methods such as k-size decomposition \cite{ref7}, k-core decomposition \cite{ref14}, k-brace decomposition \cite{ref15}, and k-truss decomposition \cite{ref16} have optimized the sizes of connected components, node degrees, and edge embedding degrees, respectively. In directed graph analysis, weak connectivity and strong connectivity methods \cite{ref12} capture community structures by either ignoring or emphasizing edge directionality, while the k-clip decomposition method \cite{ref12} refines the measurement of structural diversity by controlling the node outdegree. In terms of parameter optimization, the introduction of parameter-free structural diversity metrics breaks the limitations of fixed thresholds \cite{ref17}, thus enhancing the flexibility and adaptability of the model.

\subsection*{Literature Evaluation Metrics}
Literature evaluation metrics, essential tools for measuring academic impact, have received significant attention and undergone extensive research in recent years. Citation count is the most commonly used metric for literature evaluation, reflecting the direct academic impact of a publication \cite{ref18}. However, due to variations in citation cultures across disciplines, a single citation count cannot fully capture the structural role of a publication within the academic dissemination network. To address this limitation, the Hirsch Index (h-index) \cite{ref19} combines the number of publications and citation counts to offer a more comprehensive assessment of a scholar's academic influence. Despite this, the h-index has limitations in interdisciplinary research and across different fields. To overcome these shortcomings, researchers have proposed enhanced metrics, such as the Field-Weighted Citation Impact (FWCI) \cite{ref20}. However, traditional evaluation metrics predominantly focus on the direct impact of publications and fail to fully reveal the structural role of literature within citation networks.

To better exploit the potential of citation networks, researchers have turned to citation network-based evaluation methods. For example, degree centrality \cite{ref21} and betweenness centrality \cite{ref22}, drawn from social network analysis, assess the influence of literature by evaluating the number of neighboring nodes and the role of a node as an intermediary. Degree centrality measures the direct connections between a piece of literature and others, offering simplicity but neglecting the relative importance of neighboring nodes. In contrast, betweenness centrality highlights the role of literature as a bridge between different fields, though its calculation is complex, and its applicability to large-scale networks is limited. Furthermore, PageRank-based methods have been widely applied in literature evaluation. For instance, algorithms like SCEASRank \cite{ref23} and LeaderRank \cite{ref24} enhance the traditional PageRank algorithm by incorporating time factors, reducing bias toward older citations and placing more emphasis on the contributions of newly published literature. Additionally, the ScholarRank \cite{ref25} algorithm introduces virtual nodes to effectively address issues such as dangling nodes and strong connectivity in citation networks, improving the accuracy of literature impact evaluation. However, these methods rely on complete citation networks and are sensitive to the quality of the network structure. For example, when citation data is noisy or incomplete, the results of these methods may become unstable.

In recent years, graph embedding techniques have become a prominent focus in literature evaluation. Cunningham et al. \cite{ref26} utilized graph embedding methods to map nodes in citation networks to low-dimensional spaces, capturing complex interdisciplinary connection patterns. However, these methods have not fully exploited the structural characteristics of citation networks, especially in scenarios with limited data or incomplete networks, where embedding results may not accurately reflect node importance. Furthermore, the integration of recommendation systems with semantic analysis technologies has brought new advancements to literature evaluation. For instance, interdisciplinary literature recommendation systems that combine collaborative filtering with content embedding techniques \cite{ref27} excel at fostering cross-domain knowledge discovery and collaboration, though they still face challenges in extracting deeper semantic information.

At the same time, semantic analysis, as a crucial tool for interpreting the content of literature, has become increasingly important in literature evaluation. Pre-trained models such as BERT \cite{ref28} and SciBERT \cite{ref29} have enhanced the semantic understanding of academic texts, offering innovative approaches to literature evaluation. Specifically, SciBERT, trained on a corpus of scientific literature, more accurately captures the semantic features of academic texts. However, traditional semantic models still fail to integrate the topological characteristics of citation networks. To address this gap, this study adopts the Specter model \cite{ref30}, which enhances the understanding of both semantic and network features of literature by integrating contextual information from publication content and citation networks.

Despite expanding the research dimensions at various levels, these methods primarily rely on complete citation networks or high-quality network data, making them less adaptable to scenarios with incomplete network structures or limited data. Additionally, they often overlook the multi-dimensional attributes of literature, especially in terms of mining the semantic features and combinatorial structures of publications. To address these limitations, this study proposes a novel approach to literature evaluation. The uniqueness of this approach lies in its ability to perform effective evaluation using only reference-side data, eliminating the need for a complete citation network while incorporating the combinatorial structure and semantic information of literature.
\section*{Methods}

 The proposed model is based on the following intuition: if a paper cites multiple references from different disciplines, it reflects the researcher's integration of multidisciplinary knowledge, which contributes to forming new perspectives for interdisciplinary research and innovative research outcomes. On the other hand, a paper involving interdisciplinary studies potentially exposes researchers to a richer variety of academic perspectives and research findings. 
 
\begin{figure}[h] 
	\centering
	\includegraphics[width=0.8\textwidth]{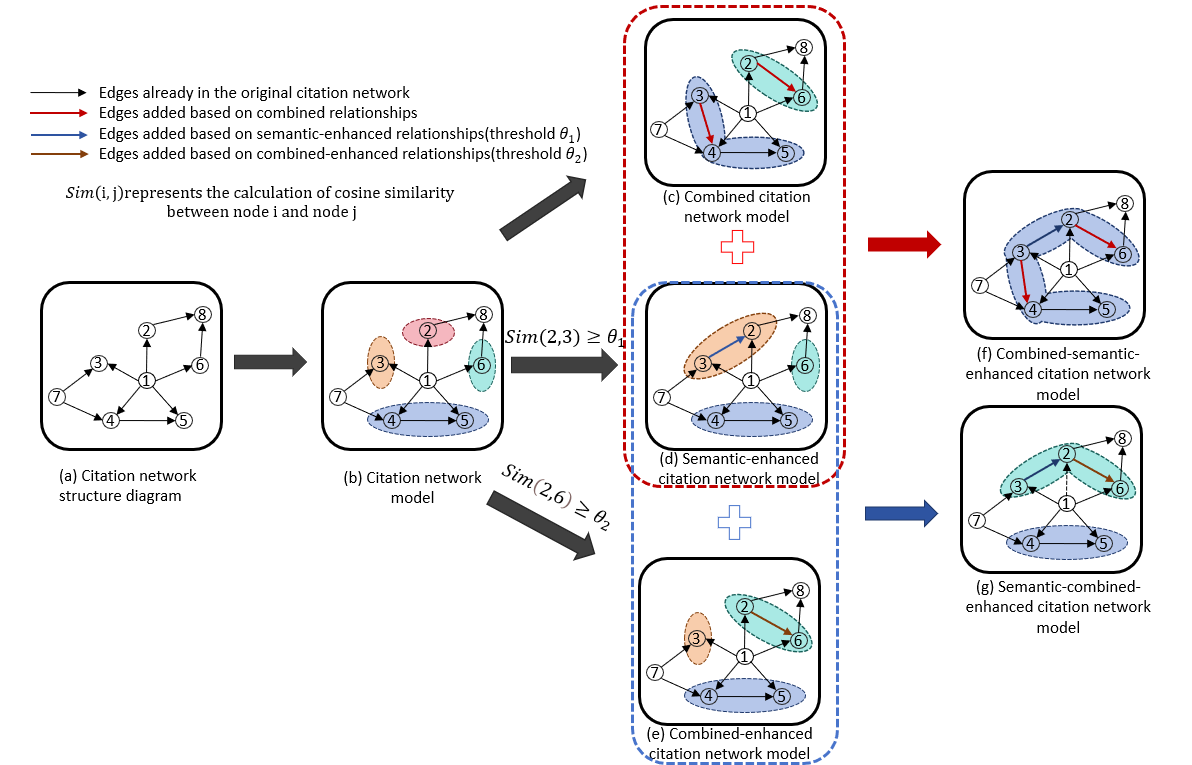}  %
	\caption{Schematic Diagram of Citation Network Enhancement Model} 
	\label{fig:citation_network} 
\end{figure}

 In this context, we introduce innovative and concise models for assessing citation structural diversity, offering fresh theoretical insights and methodologies for evaluating academic literature. The analytical procedure is depicted in Figure \ref{fig:citation_network}. Specifically, Figure \ref{fig:citation_network}(a) illustrates a citation network structure comprising nodes 1 through 8, where node 1 signifies the target literature, and nodes 2 through 6 represent its references. Node 7 cites references 3 and 4, while node 8 cites references 2 and 6.In Figure \ref{fig:citation_network}(b), the citation network model of the target literature revolves around node 1 (target literature) and its adjacent nodes, with the references of the target literature forming a subgraph. Within this subgraph, citation structural diversity is quantified by calculating the number of connected components, thereby assessing the target literature's capacity for information access and dissemination.Building upon this foundation, our study incorporates combinatorial relationships between literatures, such as coupling and co-citation, by introducing combinatorial edges. This results in the combined citation network model shown in Figure \ref{fig:citation_network}(c). Furthermore, Figure \ref{fig:citation_network}(d) presents a semantic-enhanced citation network model, incorporating semantic-enhanced edges based on a semantic similarity threshold between literatures.To further enrich the network structure, combinatorial enhancement edges are added based on a specific threshold, leading to the development of the combined-enhanced citation network model illustrated in Figure \ref{fig:citation_network}(e). Ultimately, by integrating the combined citation network model with the semantic-enhanced citation network model, we form a combined-semantic-enhanced citation network model, as depicted in Figure \ref{fig:citation_network}(f). Similarly, merging the semantic-enhanced and combined-enhanced citation network models yields the semantic-combined-enhanced citation network model in Figure \ref{fig:citation_network}(g).Through the construction of this series of models, this paper aims to enhance the application potential of citation networks in evaluating literature quality by considering citation structural features, semantic relationships, and combinatorial connections.

\subsection*{Basic Definitions}
\noindent \textbf{Definition 1:} \textbf{Citation Network Graph.} A citation network graph is denoted as \( D = \langle V, \vec{E} \rangle \), where \( V \) is the set of literatures and \( \vec{E} \) is the set of edges representing citation relationships. If a literature \( v \) references a literature \( u \), then there exists a directed edge \( \langle v, u \rangle \in \vec{E} \).

\noindent \textbf{Definition 2:} \textbf{Reference Set.} In a citation network graph \( D = \langle V, \vec{E} \rangle \), the reference set \( R_v \) of a literature \( v \in V \) is the set of all literatures cited by literature \( v \), that is, 
\begin{equation}
	R_v = \{ u \in V \mid \langle v, u \rangle \in \vec{E} \}.
\end{equation}

\noindent \textbf{Definition 3:} \textbf{Derived Subgraph of the Reference Set.} Given a citation network graph \( D = \langle V, \vec{E} \rangle \) and a reference set \( R_v \) of literature \( v \), its derived subgraph \( D[R_v] = \langle R_v, \vec{E'} \rangle \) is the subgraph consisting of the literatures in \( R_v \), where
\begin{equation} \label{eq:derived_subgraph}
	\vec{E'} = \{ \langle u, w \rangle \mid u, w \in R_v, \langle u, w \rangle \in \vec{E} \}.
\end{equation}

\noindent \textbf{Definition 4:} \textbf{Base Graph.} The base graph \( G(D) = (R_v, E') \) is the undirected graph obtained by replacing all directed edges in the graph \( D[R_v] \) with undirected edges. The set of undirected edges \( E' \) is defined as 
\begin{equation} \label{eq:base_graph}
	E' = \{ (u, w) \mid (w, u) \in \vec{E'} \vee (u, w) \in \vec{E'} \}.
\end{equation}

\noindent \textbf{Definition 5:} \textbf{Number of Connected Components.} \( |CC(G)| \) denotes the number of connected components in an undirected graph \( G = (R_v, E') \).

\subsection*{Citation Structural Diversity}

\noindent \textbf{Definition 6 Citation Structural Diversity.} 
Given a citation network graph \( D = \langle V, \vec{E} \rangle \), the derived subgraph of the reference set \( R_v \) of literature \( v \) is \( D[R_v] = \langle R_v, (\vec{E}') \rangle \). The citation structural diversity \( sd_r(v) \) of literature \( v \) is the number of connected components in the base graph of \( D[R_v] \), that is, 
\begin{equation} 
	sd_r(v) = |CC(G(D[R_v]))|.
\end{equation}

\begin{figure}[h]
	\centering
	\includegraphics[width=0.8\textwidth]{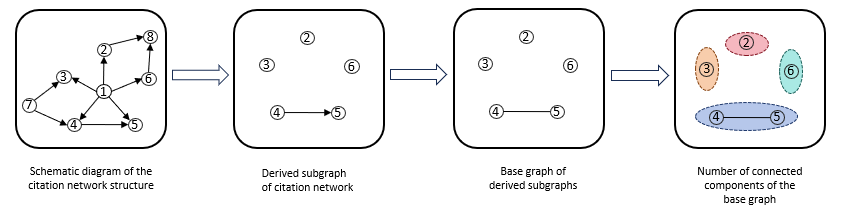}
	\caption{Schematic Diagram of Citation Structural Diversity Calculation}
	\label{fig:structural_diversity}
\end{figure}

As illustrated in the citation network model in  Figure~\ref{fig:citation_network}(b), the reference set of node 1 includes nodes 2 through 6. we derive the subgraph \( D[R_1] = \langle R_1, (\vec{E}') \rangle \), where \( R_1 = \{2, 3, 4, 5, 6\} \). In the base graph of the directed graph \( D[R_1] \), the sets \( \{2\} \), \( \{3\} \), and \( \{6\} \) each form their own connected component, while \( \{4, 5\} \) together form another connected component. Therefore, the citation structural diversity \( sd_r(1) \) of node 1 is four.Figure~\ref{fig:structural_diversity} further illustrates the calculation of citation structural diversity for literature 1 by extracting the citation network subgraph based on its reference list. The base graph of this subgraph reveals four connected components, reinforcing the citation structural diversity of literature 1 as 4.

It should be noted that all models in this paper introduce directed edges based on citation relationships, which are used solely to determine the directional relationship between citing and cited literatures. Consequently, when analyzed, connectivity in a citation network may exhibit weak connectivity rather than strong connectivity. By using the base graph \( G(D) \) of the citation network, the citation structural diversity of a literature can be measured more accurately.

\subsection*{Combined Citation Structural Diversity}
The citation structural diversity of definition 6, which include only direct citation relationships, struggle to fully capture the diversity of structures and the potential associative features among literatures. The main limitation of traditional citation networks is that they only reflect direct citation relationships, overlooking indirect ones. For instance, in citation networks, there are two common types of literature associations: first, multiple literatures cited by the same literature form co-citation relationships \cite{ref31}; second, multiple literatures that simultaneously cite the same reference generate coupling relationships \cite{ref32}. To more comprehensively capture these potential associations among literatures, we propose a combined citation network model that integrates both coupling and co-citation relationships.

To achieve this, for a citation network graph \( D = \langle V, \vec{E} \rangle \), we first construct its co-citation and coupling relations. 

The co-citation relation of \( D \) is denoted as
\begin{equation} \label{eq:co_citation}
	\vec{E_1} = \left\{ \langle u, w \rangle \mid \exists v \in V, u, w \in R_v, \exists x \in V, x \neq v, \langle x, u \rangle \in \vec{E} \wedge \langle x, w \rangle \in \vec{E} \right\}.
\end{equation}

The coupling relation of \( D \) is denoted as
\begin{equation} \label{eq:coupling}
	\vec{E_2} = \left\{ \langle u, w \rangle \mid \exists v \in V, u, w \in R_v, \exists y \in V, \langle u, y \rangle \in \vec{E} \wedge \langle w, y \rangle \in \vec{E} \right\}.
\end{equation}

Hence, the combined citation network graph of \( D \) is denoted as \( D = \langle V, \vec{E_c} \rangle \), where \( \vec{E_c} = \vec{E} \cup \vec{E_1} \cup \vec{E_2} \).

\noindent \textbf{Definition 7 Combined Citation Structural Diversity.} The derived subgraph of the combined citation structure for literature \( v \) is \( D[C_v] = \langle R_v, (\vec{E_c}') \rangle \), where
\begin{equation} \label{eq:derived_subgraph}
	(\vec{E_c}') = \left\{ \langle u, w \rangle \mid u, w \in R_v \wedge \langle u, w \rangle \in \vec{E_c} \right\}.
\end{equation}

The citation structural diversity \( sd_c(v) \) of the combined citation network for literature \( v \) is the number of connected components in the base graph of the derived subgraph \( D[C_v] \), that is,
\begin{equation} \label{eq:sd_combined}
	sd_c(v) = |CC(G(D[C_v]))|.
\end{equation}

In the combined citation network model shown in Figure~\ref{fig:citation_network}(c), both nodes 2 and 6 cite reference 8. While there is no direct connection between nodes 2 and 6 in the original citation network model, the addition of a coupling edge effectively represents their shared citation characteristics. Nodes 3 and 4 are both cited by node 7, and a new co-citation edge is introduced between nodes 3 and 4. By combining these two relationships, we merge their results to construct a combined citation network model, adding new edges to the original network. After adding the combinatorial relationships, \( \{3, 4, 5\} \) form a new connected component, and \( \{2, 6\} \) form another. Consequently, the citation structural diversity \( sd_c(1) \) of node 1 decreases from four to two.
\subsection*{Enhanced Citation Structural Diversity}

The models of  definition 6 and 7 have significant limitations in practice, restricting the models' ability to uncover deeper patterns of academic communication. First, they rely solely on explicit citation relationships, making them difficult to capture the potential value and role of literatures that have not yet been cited (e.g., results in emerging research fields or newly published papers), thus limiting their applicability in such cases. Second, the models do not fully capture the content and thematic connections among literatures. For example, some literatures may be considered similar due to citation relationships, despite having substantially different content and research focuses. This discrepancy can lead to erroneous conclusions and reduce the reliability of citation network analysis. 
To address these issues, we propose methods to enhance citation networks and mitigate the effects of missing and redundant edges. By introducing a threshold-based enhancement method, implicit semantic relationships can be more effectively supplemented while filtering out noisy edges. This approach not only captures potential associations among literatures more comprehensively but also improves the connectivity and analytical accuracy of the citation network, offering richer perspectives and better support for literature association analysis.
\subsubsection*{Semantic Extraction and Similarity Calculation}

We used a pre-trained Specter model for semantic extraction of literature titles and abstracts. Specter is a literature embedding model specifically designed for the academic domain, generating high-dimensional vector representations that capture the semantic features of literature. The model is initialized with the parameters of the pre-trained SciBERT model and learns deeper semantic associations between literatures by leveraging their titles, abstracts, and citation information. 

In the specific implementation, the embedding vectors generated by Specter are used to quantify the similarity between literatures. Using cosine similarity, an edge is added between literatures \( C_i \) and \( C_j \) if their similarity \( \text{Sim}(C_i,C_j) \) exceeds a predefined threshold \( \theta \). The formula for literature similarity is as follows:

\begin{equation}
	\text{Sim}(C_i, C_j) = \frac{\vec{v}_{C_i} \cdot \vec{v}_{C_j}}{\|\vec{v}_{C_i}\| \|\vec{v}_{C_j}\|}
\end{equation}

\noindent where \( \vec{v}_{C_i} \) and \( \vec{v}_{C_j} \) are the embedding vectors of literature \( C_i \) and \( C_j \), respectively, \( \cdot \) denotes the dot product of the vectors, \( \|\cdot\| \) represents the second norm of  the vectors, and \( \text{Sim}(C_i, C_j) \) denotes the similarity between literatures \( C_i \) and \( C_j \). This method enables accurate quantification of the semantic relationships between literatures and facilitates the construction of richer citation networks.

To further enhance the expressive power of the citation network, we introduce a similarity threshold to enrich its structure. When the similarity between two literatures exceeds a predefined threshold \( \theta \), it indicates a potential association between them, and an edge is added to the citation network. Specifically, we define two thresholds, \( \theta_1 \) and \( \theta_2 \), for adding semantic-enhanced edges and combined-enhanced edges, respectively. The addition of semantic-enhanced edges is based on the similarity between the target literature and its references, with the mean similarity value serving as the threshold \( \theta_1 \). In contrast, the addition of combined-enhanced edges is determined by the similarity between references, using the mean of the similarity values between the upper and lower quartiles (IQRMean) as the threshold \( \theta_2 \). The IQRMean is calculated by sorting the similarity values of the references and computing the mean of the values between the upper and lower quartiles, which then serves as the criterion for adding combinatorial edges. The formula for IQRMean is as follows:

\begin{equation}
	\text{IQRMean} = \frac{\sum_{i=q_1}^{q_3} x_i}{n_{q_1:q_3}}
\end{equation}

\noindent where \( q_1 \) and \( q_3 \) denote the first and third quartiles, respectively, \( x_i \) denotes the sorted similarity value, and \( n_{q_1:q_3} \) is the number of similarity values between the upper and lower quartiles. Through this dynamic adjustment of the thresholds, the structural expression of the citation network can be flexibly enhanced to provide more comprehensive support for literature relationship analysis.
\subsubsection*{Semantic-Enhanced Citation Structural Diversity}

Given a citation network graph \( D = \langle V, \vec{E} \rangle \), and with the goal of enhancing the citation network model using semantic relations, we first construct the semantic relations. When the semantic similarity between two literatures exceeds a threshold \( \theta_1 \), a semantic-enhanced edge is added to the citation network. The semantic relation of \( D \) is denoted as \(\vec{E_3} = \left\{ \langle u, w \rangle \mid \exists v \in V, u, w \in R_v, \langle u, w \rangle \notin \vec{E} \wedge \text{Sim}(u, w) \geq \theta_1 \right\}\).Thus, the semantic-enhanced citation network graph of \( D \) is denoted as \( D = \langle V, (\vec{E}_{ss}) \rangle \), where \( (\vec{E}_{ss}) = \vec{E} \cup \vec{E}_3) \).

\noindent \textbf{Definition 8:} \textbf{Semantic-Enhanced Citation Structural Diversity.} The derived subgraph of the semantic-enhanced citation network for literature \( v \) is \( D[SS_{(v, \theta_1)}] = \langle R_v, (\vec{E'}_{ss}) \rangle \), where \( (\vec{E'}_{ss}) = \{ \langle u, w \rangle \mid u, w \in R_v \land \langle u, w \rangle \in (\vec{E}_{ss}) \} \). The semantic-enhanced citation structural diversity \( \text{sd}_{ss}(v) \) of literature \( v \) is the number of connected components in the base graph of the derived subgraph \( D[SS_{(v, \theta_1)}] \), that is, 

\begin{equation}
	\text{sd}_{ss}(v) = |CC(G(D[SS_{(v, \theta_1)}]))|
\end{equation}

In the semantic-enhanced citation network model presented in Figure~\ref{fig:citation_network}(d), semantic-enhanced edges are added between nodes 2 and 3 when the semantic similarity \( \text{Sim}(2, 3) \) between node 2 and node 3, which originally lacked direct connectivity, satisfies \( \text{Sim}(2, 3) \geq \theta_1 \). This enhancement method significantly improves the connectivity and expressiveness of the network. In the figure, the citation structural diversity \( \text{sd}_{ss}(1) \) of node 1 is initially four, but after adding semantic-enhanced edges, nodes 2 and 3 merge into one connected component, reducing the citation structural diversity \( \text{sd}_{ss}(1) \) to three.
\subsubsection*{Combined-Enhanced Citation Structural Diversity}

The model combines coupling relations and co-citation relations and sets a similarity threshold \( \theta_2 \) for combinatorial relations as a restriction. Unlike the traditional combined citation network model, the combined-enhanced citation network model filters the combined relations and adds edges to the citation network only when the similarity threshold \( \theta_2 \) is satisfied.

Given a citation network graph \( D = \langle V, \vec{E} \rangle \), the construction of a combinatorially enhanced citation network requires the establishment of its coupling enhancement relation and co-citation enhancement relation. 

The co-citation enhancement relation is expressed as \( \vec{E}_4 \), where 

\begin{equation}
	\vec{E_4} = \left\{ \langle u,w \rangle \mid \exists v \in V,u, w \in R_v, \exists x \in V, x \neq v, \left( \langle x, u \rangle \in \vec{E} \wedge \langle x, w \rangle \in \vec{E} \right) \wedge \left( \langle u, w \rangle \notin \vec{E} \wedge \text{Sim}(u, w) \geq \theta_2 \right) \vee \left( \langle w, u \rangle \notin \vec{E} \wedge \text{Sim}(w, u) \geq \theta_2 \right) \right\}.
\end{equation}

The coupling enhancement relation is expressed as \( \vec{E}_5 \), where
\begin{equation}
	\vec{E_5} = \left\{ \langle u, w \rangle \mid \exists v \in V, u, w \in R_v, \exists y \in V, \left( \langle u, y \rangle \in \vec{E} \wedge \langle w, y \rangle \in \vec{E} \right) \wedge \\\left( \left( \langle u, w \rangle \notin \vec{E} \wedge \text{Sim}(u, w) \geq \theta_2 \right) \vee \right. \left. \left( \langle w, u \rangle \notin \vec{E} \wedge \text{Sim}(w, u) \geq \theta_2 \right) \right) \right\}.
\end{equation}

Thus, the combined-enhanced citation network \( D \) is denoted as \( D = \langle V, (\vec{E}_{cs}) \rangle \), where
\begin{equation}
	(\vec{E}_{cs}) = \vec{E} \cup \vec{E}_4 \cup \vec{E}_5
\end{equation}
\noindent \textbf{Definition 9:} \textbf{Combined-Enhanced Citation Structural Diversity.} The derived subgraph of the combined-enhanced citation structure for literature \( v \) is \( D[CS_{(v, \theta_2)}] = \langle R_v, (\vec{E'}_{cs}) \rangle \), where 
\begin{equation}
	(\vec{E'}_{cs}) = \{ \langle u, w \rangle \mid u, w \in R_v \land \langle u, w \rangle \in (\vec{E}_{cs}) \}
\end{equation}
The combined-enhanced citation structural diversity \( \text{sd}_{cs}(v) \) of literature \( v \) is the number of connected components in the base graph of the derived subgraph \( D[CS_{(v, \theta_2)}] \), that is,
\begin{equation}
	\text{sd}_{cs}(v) = |CC(G(D[CS_{(v, \theta_2)}]))|
\end{equation}
In the combined-enhanced citation network model shown in Figure~\ref{fig:citation_network}(e), nodes 2 and 6 both cite node 8, while nodes 3 and 4 are cited by the same node 7. In the combined-enhanced citation network model, we introduce a similarity threshold \( \theta_2 \). For example, if nodes 3 and 4 have a co-citation relationship but \( \text{Sim}(3, 4) < \theta_2 \), no co-citation enhancement edges are added; whereas if the similarity between nodes 2 and 6 exceeds the threshold, that is, \( \text{Sim}(2, 6) \geq \theta_2 \), which satisfies the condition for combinatorial enhancement, a coupling enhancement edge is added between them. When calculating citation structural diversity, \( \{ 2, 3 \} \) form a connected component, \( \{ 4, 5 \} \) form another connected component, and \( \{ 6 \} \) becomes a separate connected component. Therefore, in this enhanced network, there are three connected components and the citation structural diversity \( \text{sd}_{cs}(1) \) of node 1 is three.
\subsubsection*{Semantic-Combined-Enhanced Citation Structural Diversity}

The model combines semantic similarity between literatures with combinatorial relationships and sets different similarity thresholds \( \theta_1 \) and \( \theta_2 \), resulting in a more comprehensive and complex network structure.

\noindent \textbf{Definition 10:} \textbf{Semantic-Combined-Enhanced Citation Structural Diversity.} Given a citation network graph \( D = \langle V, \vec{E} \rangle \), the derived subgraph of the semantic-combined-enhanced citation network structure for literature \( v \) is 

\begin{equation}
	D[SCS_v] = D[SS_{(v, \theta_1)}] \cup D[CS_{(v, \theta_2)}]
\end{equation}

The semantic combination of literatures \( v \) enhances the citation structural diversity of the citation network. The citation structural diversity \( \text{sd}_{scs}(v) \) is the number of connected components in the base graph of the derived subgraph \( D[SCS_v] \), that is,

\begin{equation}
	\text{sd}_{scs}(v) = |CC(G(D[SCS_v]))|
\end{equation}

In a specific application, the semantic combination relation first considers the semantic similarity between the literatures. When the semantic similarity between node 2 and node 3 exceeds a set threshold, that is, \( \text{Sim}(2, 3) \geq \theta_1 \), the model will add an edge between these two nodes. In addition, the introduction of semantic combination relation allows the model to more accurately describe the complex connections between literatures. For example, although there is a co-citation relationship between node 3 and node 4 in the semantic-combined-enhanced citation network model of Figure~\ref{fig:citation_network}(g), node 3 and node 4 will not have an edge added between them if their similarity does not reach a set threshold, that is, \( \text{Sim}(3, 4) < \theta_2 \). On the other hand, the similarity between node 2 and node 6 reaches the predetermined threshold \( \text{Sim}(2, 6) \geq \theta_2 \), so the model will add edges between them. Thus, \( \{2, 3, 6\} \) forms a connected component, and \( \{4, 5\} \) forms another connected component. Therefore, the citation structural diversity \( \text{sd}_{scs}(1) \) of node 1 is two.
\subsubsection*{Combined-Semantic-Enhanced Citation Structure Diversity}

In this model, we combine semantic relations with traditional combinatorial relations. Unlike the combined-enhanced citation network model, we do not impose a similarity threshold when adding combined relationships. That is, as long as a coupling or co-citation relation exists, the corresponding edge can be added directly to the network without considering additional semantic similarity requirements. However, when adding semantic relations, the model still filters them by setting the threshold \( \theta_1 \).

\noindent \textbf{Definition 11:} \textbf{Combined-Semantic-Enhanced Citation Structure Diversity.} Given a citation network graph \( D = \langle V, \vec{E} \rangle \), the combined-semantic-enhanced citation network structure derived subgraph of literature \( v \) is 

\begin{equation}
	D[CSS_v] = D[C_v] \cup D[SS_{(v, \theta_1)}]
\end{equation}

The combinatorial semantics of literature \( v \) enhances the structural diversity of the citation network. The citation structural diversity \( \text{sd}_{css}(v) \) is the number of connected components in the base graph of the derived subgraph \( D[CSS_v] \), that is,

\begin{equation}
	\text{sd}_{css}(v) = |CC(G(D[CSS_v]))|
\end{equation}

In this model, for example, in the combined-semantic-enhanced citation network model shown in Figure~\ref{fig:citation_network}(f), node 2 and node 6 co-cite a reference, while node 3 and node 4 share a co-citation relationship. Based on these citation behaviors, the model directly adds edges reflecting these relationships to the citation network model without applying any threshold. Meanwhile, for node pairs with semantic similarity, such as nodes 2 and 3, when their semantic similarity exceeds the set threshold \( \theta_1 \), that is, if \( \text{Sim}(2, 3) \geq \theta_1 \), an edge is added based on the semantic similarity between the two nodes. As a result, the connected components in the graph are \( \{2, 3, 4, 5, 6\} \), and the corresponding citation structural diversity metric \( \text{sd}_{css}(1) \) is calculated as one.
\section*{Experimentation}
This experiment aims to investigate the relationship between the citation structural diversity of literature and citation volume, while enhancing the accuracy of citation prediction models through the optimization of citation networks. Initially, the experiment utilizes the DBLP Citation Network V13 dataset from the Aminer platform to examine how citation structural diversity impacts citation volume. Additionally, the experiment further confirms the significant correlation between citation structural diversity and citation volume through long-term citation analysis, evaluation of citation prediction models, and comparison of various semantic models. To explore the interdisciplinarity of literature, the experiment also incorporates the PubMed dataset, combining citation structural diversity analysis to examine the interdisciplinary characteristics of the literature.

\subsection*{Data Preparation and Processing}
We used the DBLP Citation Network V13 dataset provided by the Aminer platform \cite{ref33}. However, some literature entries had missing or insufficient citations, which affected the completeness of the citation network and the accuracy of the citation structural diversity measurement. Additionally, some bibliographic information (e.g., author names, titles, or publication details) was incomplete, which hindered literature association analysis and evaluation. To address these issues, we performed data cleaning prior to usage and selected the largest connected component for analysis to ensure the rigor and representativeness of the experiment.

During the data preprocessing stage, we first categorized the literature by year and publication type (journals, conferences, books, etc.). We then selected publications from key years: 2000, 2006, 2012, and 2018. These key years were chosen because literature published before 2000 is relatively old, and there are fewer publications after 2018. These years better reflect the evolving trends and cyclical characteristics of the literature. Although the literature was categorized by year and publication type, variations in publication quality still exist. To ensure the rigor of the experiment, we further graded the journals and conferences according to internationally recognized classification standards. Journals were classified according to the Journal Citation Reports (JCR) standard into four categories: Q1, Q2, Q3, and Q4. Conferences were classified based on the Chinese Computer \textnormal{Society's} Recommended International Academic Conference and Journal Catalogue (CCF) standard into three categories: A, B, and C.

To better present the experimental results, we selected seven data sets based on publication categories from the classified data. Specific information is shown in Table \ref{tab:literature_info}.

\begin{table}[ht]
	\centering
	
	\normalsize 
	\begin{tabular}{|l|l|l|l|l|}  
		\hline
		& \textbf{Name of Publication} & \textbf{Ranking of Publication} & \textbf{Year} & \textbf{Number of literatures} \\ 
		\hline
		Group 1 & IEEE Transactions on Communications & Q1& 2000  & 250 \\ 
		\hline
		Group 2 & Journal of Computational Chemistry  & Q2 & 2018& 301 \\ 
		\hline
		Group 3 & Discrete Applied Mathematics  & Q3& 2006 & 202 \\ 
		\hline
		Group 4 & Journal of Computer Science and Technology  & Q4& 2006 & 106 \\ 
		\hline
		Group 5 & CHI  & A& 2012 & 348 \\ 
		\hline
		Group 6 & ICDCS  & B & 2006& 71 \\ 
		\hline
		Group 7 & ICTAI  & C& 2018 & 146 \\ 
		\hline
	\end{tabular}
	\caption{Specific Information on Seven Groups of Literature from the Same Year and Journal}
	\label{tab:literature_info}
\end{table}
\subsection*{Experiments on the Correlation Between Citation Structural Diversity and Citation Volume}

As a core indicator of academic impact, citation volume reflects the academic value and recognition of a work to some extent \cite{ref4}. Highly cited literature is typically a significant research contribution in related fields and is widely referenced by scholars. Therefore, citation volume plays a crucial role in assessing the quality and academic recognition of literature. Investigating the relationship between citation structural diversity and citation volume helps reveal how citation network structural features influence academic impact, providing a theoretical foundation for optimizing citation network models.

To explore the impact of citation structural diversity on literature quality, we examine its correlation with citation volume. First, a citation network is constructed based on citation relationships among the literature. Next, the Specter model and tokenizer are loaded to link the titles and abstracts of the literatures into semantically complete texts. These texts are then tokenized and encoded, and the [CLS] token embedding corresponding to each literature is extracted as its semantic representation. Finally, semantic similarity between literatures is quantified by calculating cosine similarity, which helps enhance and optimize the citation network.

After enhancing the citation network, we calculate the citation structural diversity of each literature based on the optimized network structure and analyze it alongside citation statistics over a three-year period. To investigate the relationship between citation structural diversity and citation volume in more detail, we group the literature based on citation structural diversity values and use two statistical metrics to characterize the citation distribution. The first metric is the median citation, which robustly reflects the central tendency of the citation distribution and minimizes the influence of extreme values. The second metric is the IQRMean of citations, which measures the concentration and overall trend of citation volume. These two metrics complement each other, providing a reliable basis for understanding citation patterns in different citation structural diversity groupings.

To quantify the correlation between citation structural diversity and citation volume, we calculate the Pearson correlation coefficient, denoted as \( r \), between the two variables. The formula is as follows:

\begin{equation}
	r = \frac{\sum_{i} (x_i - \bar{x}) (y_i - \bar{y})}{\sqrt{\sum_{i} (x_i - \bar{x})^2 \sum_{i} (y_i - \bar{y})^2}}
	\quad 
\end{equation}

\noindent where $x_i$ and $y_i$ represent the citation structural diversity value and the number of citations for the $i$-th literature, respectively, and $\bar{x}$ and $\bar{y}$ are their mean values. The Pearson correlation coefficient $r$ ranges from -1 to 1, where values closer to 1 indicate a stronger positive correlation, and values closer to -1 indicate a stronger negative correlation. The results of the Pearson correlation analysis based on the two statistical metrics (median citations and IQRMean of citations) are presented in Tables \ref{tab:median_citation_correlation} and \ref{tab:iqrmean_citation_correlation}.

\begin{table}[ht]
	\centering
	
	\resizebox{\textwidth}{!}{
		\begin{tabular}{|>{\raggedright\arraybackslash}p{1.5cm}| 
				>{\raggedright\arraybackslash}p{2.0cm}| 
				>{\raggedright\arraybackslash}p{3.0cm}| 
				>{\raggedright\arraybackslash}p{3.5cm}| 
				>{\raggedright\arraybackslash}p{3.5cm}| 
				>{\raggedright\arraybackslash}p{3.5cm}| 
				>{\raggedright\arraybackslash}p{3.5cm}|}
			\hline
			\textbf{Group} & 
			\textbf{Citation Network Model} & 
			\textbf{Combined Citation Network Model} & 
			\textbf{Semantic-Enhanced Citation Network Model} & 
			\textbf{Combined-Enhanced Citation Network Model} & 
			\textbf{Semantic-Combined-Enhanced Citation Network Model} & 
			\textbf{Combined-Semantic-Enhanced Citation Network Model} \\
			\hline
			Group 1 & 0.531 & 0.142 & \textbf{\textit{0.811}} & 0.589 & 0.783 & 0.188 \\
			\hline
			Group 2 & 0.545 & 0.730 & 0.828 & 0.565 & 0.828 & \textbf{\textit{0.924}} \\
			\hline
			Group 3 & 0.489 & 0.626 &
			 \textbf{\textit{0.637}} & 0.480 & \textbf{\textit{0.637}} & 0.629 \\
			 \hline
			Group 4 & 0.234 & 0.086 & \textbf{\textit{0.504}} & 0.286 & 0.421 & 0.273 \\
			\hline
			Group 5 & 0.429 & 0.226 & 0.526 & 0.588 & \textbf{\textit{0.798}} & 0.639 \\
			\hline
			Group 6 & 0.416 & 0.593 & 0.652 & 0.532 & 0.649 & \textbf{\textit{0.866}} \\
			\hline
			Group 7 & 0.346 & 0.411 & 0.659 & 0.355 & 0.649 & \textbf{\textit{0.713}} \\
			\hline
			\textbf{Mean} & 0.427 & 0.402 & 0.659 & 0.485 & \textbf{\textit{0.681}} & 0.605 \\
			\hline
			\textbf{Variance} & \textbf{\textit{0.010}} & 0.056 & 0.013 & 0.012 & 0.017 & 0.067 \\
			\hline
		\end{tabular}
	}
	\caption{Pearson Correlation Results Based on Median Citations}
	\label{tab:median_citation_correlation}
\end{table}

\begin{table}[ht]
	\centering
	
	\resizebox{\textwidth}{!}{
	\begin{tabular}{|>{\raggedright\arraybackslash}p{1.5cm}| 
			>{\raggedright\arraybackslash}p{2.0cm}| 
			>{\raggedright\arraybackslash}p{3.0cm}| 
			>{\raggedright\arraybackslash}p{3.5cm}| 
			>{\raggedright\arraybackslash}p{3.5cm}| 
			>{\raggedright\arraybackslash}p{3.5cm}| 
			>{\raggedright\arraybackslash}p{3.5cm}|} 
		\hline
		\textbf{Group} & 
		\textbf{Citation Network Model} & 
		\textbf{Combined Citation Network Model} & 
		\textbf{Semantic-Enhanced Citation Network Model} & 
		\textbf{Combined-Enhanced Citation Network Model} & 
		\textbf{Semantic-Combined-Enhanced Citation Network Model} & 
		\textbf{Combined-Semantic-Enhanced Citation Network Model} \\
		\hline
		Group 1 & 0.689 & 0.324 &\textbf{\textit{0.851}} & 0.698 & 0.786 & 0.437 \\
		\hline
		Group 2 & 0.482 & 0.738 & 0.919 & 0.630 & 0.922 & \textbf{\textit{0.931}} \\
		\hline
		Group 3 & 0.607 & 0.706 & 0.739 & 0.632 & \textbf{\textit{0.747}} & 0.730 \\
		\hline
		Group 4 & 0.219 & 0.136 & \textbf{\textit{0.470}} & 0.296 & 0.452 & 0.304 \\
		\hline
		Group 5 & 0.497 & 0.295 & 0.584 & 0.613 & 0.869 & \textbf{\textit{0.896}} \\
		\hline
		Group 6 & 0.445 & 0.566 & 0.637 & 0.526 & 0.642 & \textbf{\textit{0.852}} \\
		\hline
		Group 7 & 0.322 & 0.459 & 0.678 & 0.396 & 0.677 & \textbf{\textit{0.680}} \\
		\hline
		\textbf{Mean} & 0.466 & 0.461 & 0.697 & 0.542 & \textbf{\textit{0.728}} & 0.690 \\
		\hline
		\textbf{Variance} & 0.022 & 0.043 & 0.020 & \textbf{\textit{0.018}} & 0.021 & 0.049 \\
		\hline
	\end{tabular}
	}
	\caption{Pearson Correlation Results Based on Citation Volume IQRMean}
	\label{tab:iqrmean_citation_correlation}
\end{table}

The results from Tables \ref{tab:median_citation_correlation} and \ref{tab:iqrmean_citation_correlation} indicate a significant positive correlation between citation structural diversity and citation volume, suggesting that literature with higher citation structural diversity tends to receive higher citation volumes. Regardless of the statistical method used, the enhanced citation network models outperform the traditional citation network model in terms of correlation with citation volume. Among the enhanced models, the semantic-combined-enhanced citation network model shows the strongest correlation, particularly in both the median citations and IQRMean, highlighting its effectiveness in capturing the citation patterns of literature.

To more intuitively demonstrate the impact of citation structural diversity on citations, we divided the citation structural diversity into three groups: low diversity (1-3), medium diversity (4-6), and high diversity (7 and above). We then calculated the average citations for each group to analyze the relationship between citation structural diversity and citation volume. Based on this, we selected the data from Group 5 and visually compared the mean citation volume of each diversity group under different models using histograms. The bar chart distinguishes the data distribution of low, medium, and high diversity groups through color shading, clearly illustrating the specific trend of citation structural diversity's impact on citation volume.

\begin{figure}[ht]
	\centering
	\begin{subfigure}[(a)]{0.45\textwidth}
		\centering
		\includegraphics[width=\textwidth]{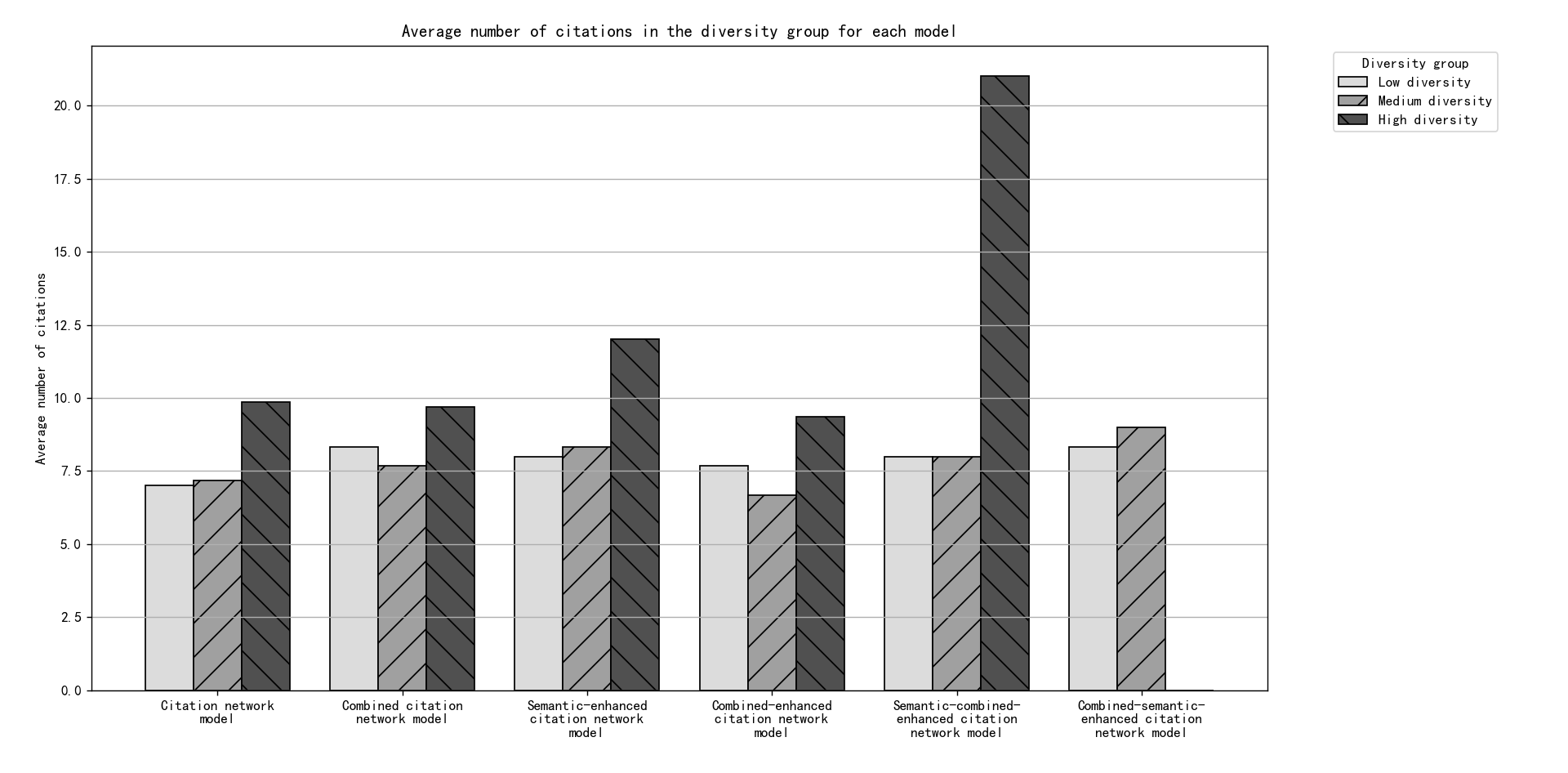}
		\caption{Plotted as Median Citations}
		\label{fig:fig3}
	\end{subfigure}
	\hfill
	\begin{subfigure}[(b)]{0.45\textwidth}
		\centering
		\includegraphics[width=\textwidth]{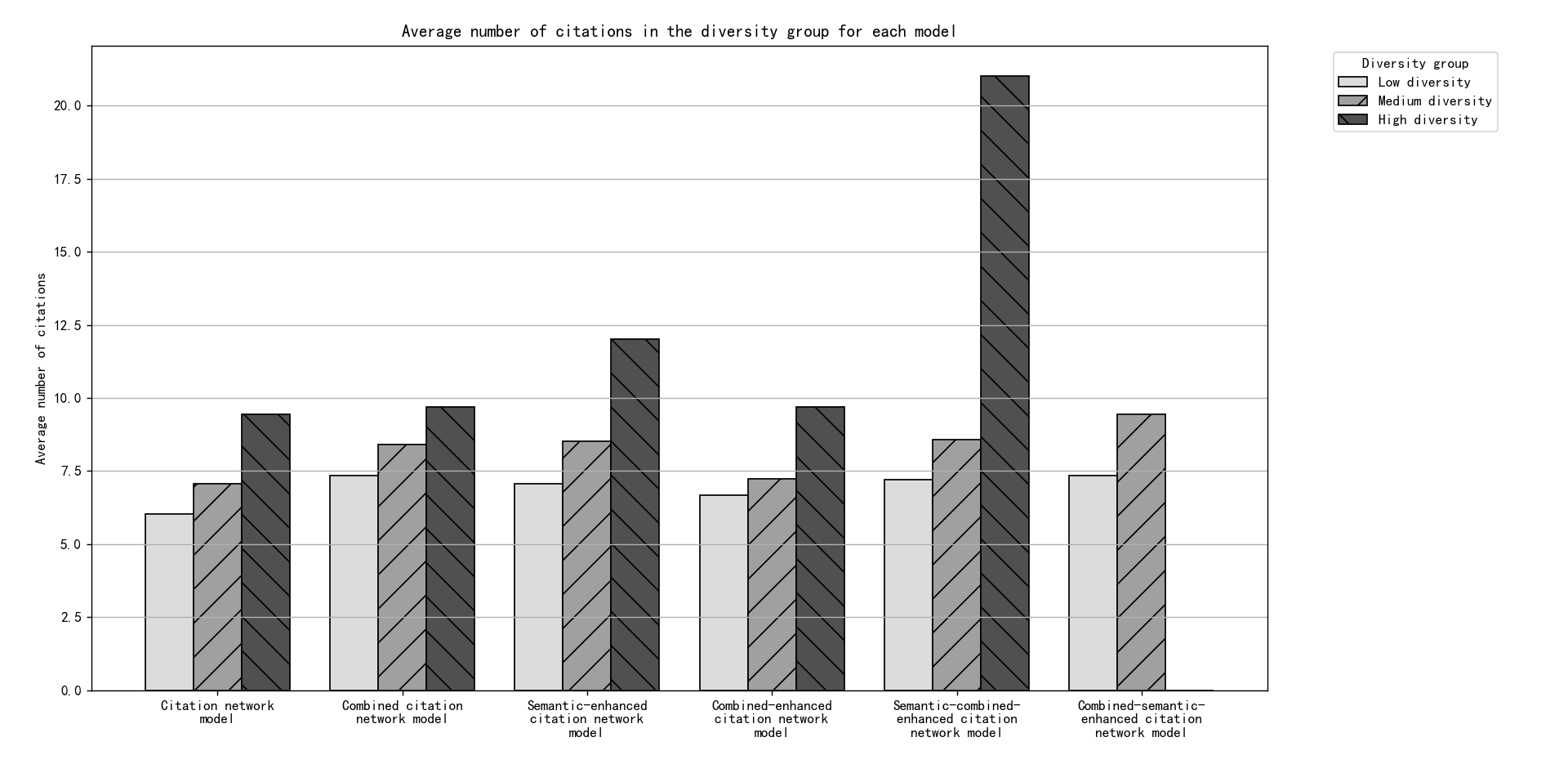}
		\caption{Plotted as IQRMean Citations}
		\label{fig:fig4}
	\end{subfigure}
	\caption{Different Group of Citation Structural Diversity vs. Citation}
	\label{fig:combined}
\end{figure}

Additionally, two metrics were used to calculate citations for each citation structural diversity group: the median of citations (see Figure \ref{fig:fig3}) and the IQRMean of citations (see Figure \ref{fig:fig4}).

By examining the histogram, we observe a significant upward trend in the number of citations as citation structural diversity increases. This suggests that literature with higher citation structural diversity generally receives more citations, supporting our experimental hypothesis to some extent. Notably, the increase in citations is most pronounced in the high diversity group, while the low diversity group shows relatively fewer citations. This further suggests that higher citation structural diversity is associated with greater citation volume, with the citation advantage being more prominent in the high diversity group and more limited in the low diversity group. 

Additionally, the semantic-combined-enhanced citation network model demonstrates clear superiority in grouping effects, further validating citation structural diversity as a key factor influencing citation volume. The results provide strong theoretical support for using citation structural diversity as an optimized metric for literature evaluation.
\subsection*{A Study of Citation Structural Diversity and Citation Trends Over a Ten-Year Period}

The goal of examining citation structural diversity and citation trends over a ten-year period is to explore the impact of citation structural diversity on long-term citation trends and analyze the potential relationship between citation structural diversity and citation volume. The operationalization of the study involved first extracting the citation structural diversity values at the reference end of each literature and categorizing them into three groups: low diversity (1-3), medium diversity (4-6), and high diversity (7 and above). Then, the number of citations for each literature within 10 years of its publication was counted. The ten-year period was chosen because a longer time window is more effective in revealing citation trends and the evolution of academic impact, while short-term data may not adequately reflect the long-term value of literature \cite{ref35}. For example, literature published later (e.g., in 2018) could not provide complete ten-year citation data. Therefore, literature published in 2000, 2006, and 2012 were selected for analysis in this study.

To eliminate absolute differences in citation volumes among different literatures, the cumulative number of citations over the ten years was calculated and normalized, making the data from different literatures more comparable. The normalized citation data help to more intuitively observe citation trends over time and enable a more accurate analysis of the correlation between citation structural diversity and citation volume. The normalization process is given by the formula:

\begin{equation}
	\text{Normalized}_i = \frac{Q_i}{\sum_{t=1}^{10} Q_t} 
\end{equation}
\noindent Where, \( \text{Normalized}_i \) denotes the normalized citations of the literature in year \( i \), \( Q_i \) denotes the actual citations of the literature in year \( i \), and \( \sum_{t=1}^{10} Q_t \) denotes the total citations of the literature in 10 years.

\begin{figure}[htbp]
	 \centering 
	\begin{minipage}[b]{0.45\linewidth}
		\centering
		\includegraphics[width=\linewidth]{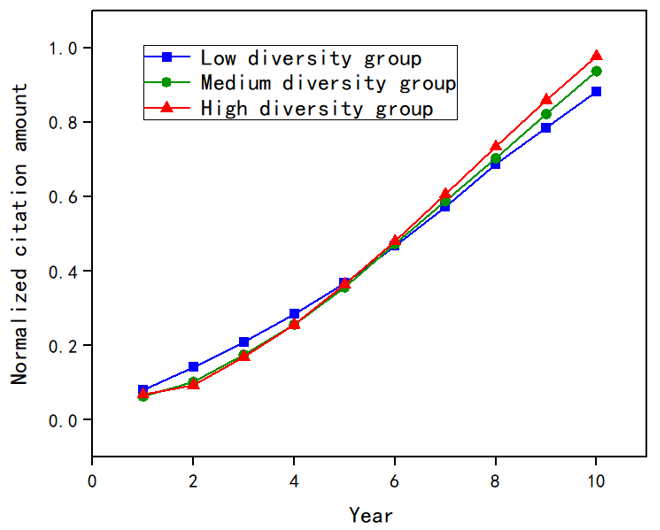}
		\subcaption{Citation Network Model} 
	\end{minipage}
	\hfill
	\begin{minipage}[b]{0.45\linewidth}
		\centering
		\includegraphics[width=\linewidth]{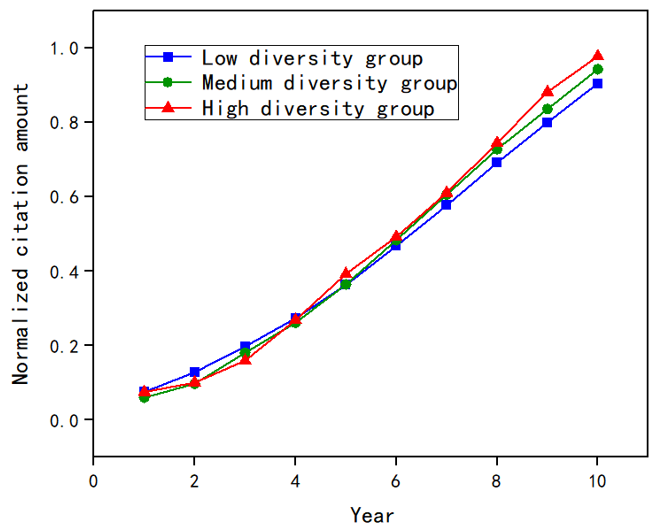}
		\subcaption{Combined Citation Network Model} 
	\end{minipage}
	
	\par\vspace{1em} 
	
	\begin{minipage}[b]{0.45\linewidth}
		\centering
		\includegraphics[width=\linewidth]{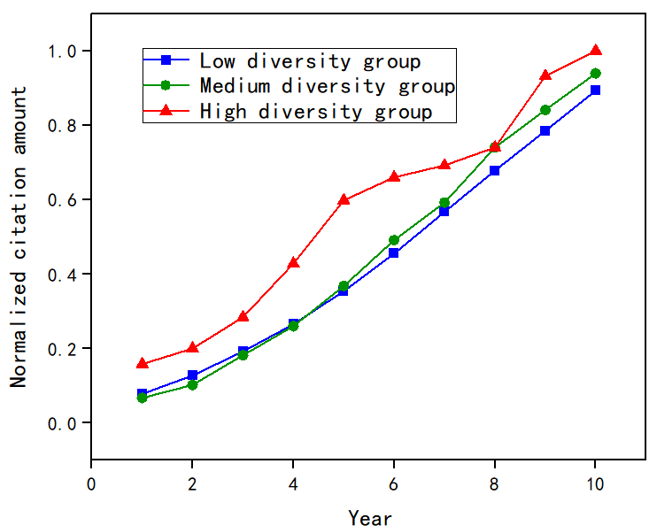}
		\subcaption{Semantic-Enhanced Citation Network Model} 
	\end{minipage}
	\hfill
	\begin{minipage}[b]{0.45\linewidth}
		\centering
		\includegraphics[width=\linewidth]{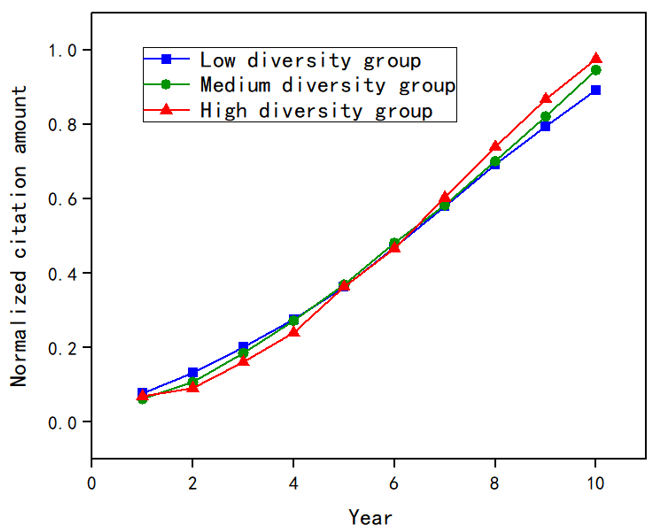}
		\subcaption{Combined-Enhanced Citation Network Model} 
	\end{minipage}
	
	\par\vspace{1em} 
	
	\begin{minipage}[b]{0.45\linewidth}
		\centering
		\includegraphics[width=\linewidth]{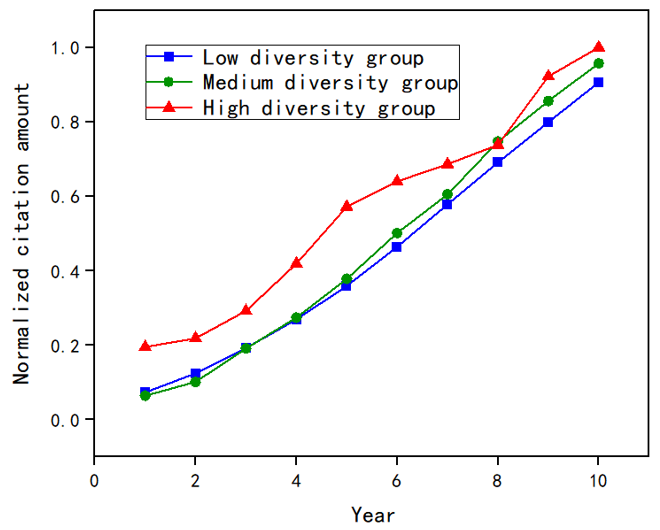}
		\subcaption{Semantic-Combined-Enhanced Citation Network Model} 
	\end{minipage}
	\hfill
	\begin{minipage}[b]{0.45\linewidth}
		\centering
		\includegraphics[width=\linewidth]{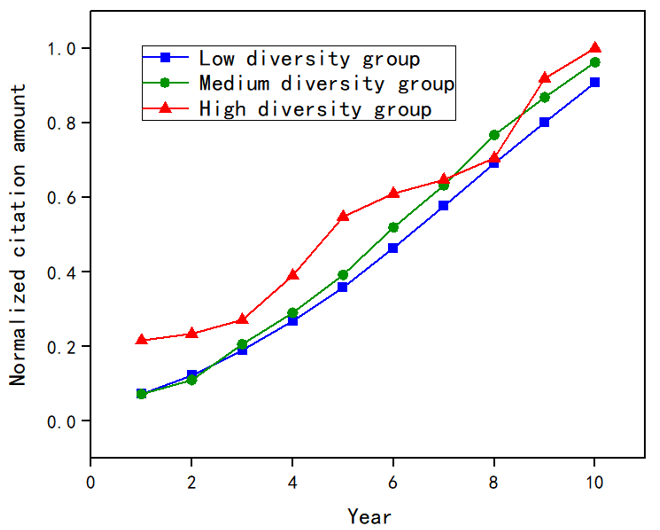}
		\subcaption{Combined-Semantic-Enhanced Citation Network Model} 
	\end{minipage}
	
	\caption{Experiment on citation structural diversity and Literature Citation Trends Over Time} 
	\label{fig:fig5}
\end{figure}

Based on the normalized citation data, we calculated the Pearson correlation coefficients for the low, medium, and high diversity groups to assess the relationship between citation trends over time at different levels of citation structural diversity. This grouping analysis not only reveals the impact of citation structural diversity on citation trends but also validates the differences in citation network model performance across the different diversity groups. To visualize the citation trends, we plotted citation trend graphs for the three diversity groups, using different colors (blue, green, and red) to represent the citation trend curves of the low, medium, and high diversity groups, respectively. This visualization allows readers to intuitively understand the impact of citation structural diversity on citation changes over time. Figure \ref{fig:fig5} shows citation trend graphs for six models, selected from literature published in 2000, 2006, and 2012. The blue color represents the low diversity group, green represents the medium diversity group, and red represents the high diversity group.

The experimental results show that the citation volume for the high diversity group is consistently significantly higher than that for the medium and low diversity groups, regardless of the model used. Specifically, in the proposed enhanced citation network model, the model is able to clearly distinguish between the high, medium, and low diversity groups in most cases, effectively highlighting the differences in citation volumes among the groups. Literature with higher citation structural diversity demonstrates a significant advantage in long-term citation growth. Although the citation advantage of the high diversity group is not apparent at the beginning of the study in some models (e.g., the combined-enhanced citation network model), its citation volume gradually surpasses that of the low and medium diversity groups over time. This suggests that literature with high diversity may not have gained widespread attention initially, but over time, its scholarly value has been increasingly recognized, reflecting the importance of a longer observation window in assessing literature's diversity and impact.

However, this study also reveals some limitations. For instance, the threshold-based edge increase strategy used in constructing the enhanced citation network model resulted in a relatively small number of literatures in the high diversity group and a larger number in the medium and low diversity groups, which may have impacted the analysis results. Additionally, some models had limited ability to distinguish differences between diversity groups in the early stages of the study, suggesting that future research should optimize model design to improve distinguishing accuracy.

Overall, the findings further validate the positive correlation between citation structural diversity and the growth of literature citations, highlighting the critical role of the time factor in revealing dynamic changes in academic impact. This provides an important reference for further exploration of the academic value of citation structural diversity and its application in citation network modeling, as well as evidence that citation structural diversity performs better in terms of long-term effects as an optimized literature evaluation index.

\subsection*{Assessment Performance in Citation Predictive Modeling}

The literature sets from the 2000, 2006, and 2012 publications were selected for this experiment, with 80\% of the dataset used as the training set and 20\% as the test set. The aim of the experiment is to evaluate and optimize the performance of the citation prediction model. Five classical regression models were chosen for the prediction analysis, including linear regression (LR) \cite{ref36}, K-nearest neighbor (KNN) \cite{ref37}, support vector regression (SVR) \cite{ref38}, and classification and regression tree (CART) \cite{ref39}. Additionally, Multilayer Perceptron (MLP) \cite{ref40} was included as a deep learning model. These models perform regression prediction by directly utilizing literature features, while citation structural diversity features at the reference end of the literature samples were added to the original models to assess the role of citation structural diversity in citation prediction.

Five classical models were used for citation prediction analysis in this experiment, instead of more advanced models that have emerged in recent years. This decision was made because the primary goal of the experiment is to evaluate the performance of citation structural diversity features in citation prediction, rather than simply comparing the predictive performance of the models. Classical models are relatively simple, easy to interpret, and can more directly demonstrate the effect of adding citation structural diversity features on prediction accuracy. This approach helps clarify the contribution of specific features without the interference of complex, "black-box" models. Furthermore, classical models are well-established and widely used, allowing researchers to easily replicate experimental results and analyze the specific impact of citation structural diversity features.

The experiments used the coefficient of determination \( R^{2} \) and mean square error (MSE) as the primary evaluation metrics. \( R^{2} \) measures the degree of fit between the model predictions and the actual results, with values closer to 1 indicating higher prediction accuracy. MSE measures the deviation between predicted and actual values, with smaller values indicating higher accuracy. The specific parameter settings are shown in Table \ref{tab:parameters}:

\begin{table}[htbp]
	\centering

	\renewcommand{\arraystretch}{1.5}  
	\normalsize
	\begin{tabular}{|>{\raggedright\arraybackslash}p{3cm} |>{\raggedright\arraybackslash}p{8cm}|}  
		\hline
		\textbf{\raggedright Algorithm name} & \textbf{\raggedright Parameter setting} \\ 
		\hline
		LR                      & n\_jobs=1, rest of the parameters are default \\ 
		\hline
		KNN                     & n\_neighbors=7, weights='uniform', algorithm='auto', leaf\_size=40, p=1, metric='minkowski' \\ 
		\hline
		SVR                     & kernel='poly', C=1000.0, degree=2 \\ 
		\hline
		CART                    & criterion='squared\_error', splitter='random', min\_samples\_split=10, min\_samples\_leaf=2, random\_state=0 \\ 
		\hline
		MLP                     & hidden\_layer\_sizes=(200,100), activation='tanh', solver='adam', learning\_rate='adaptive', learning\_rate\_init=0.01, max\_iter=1000, random\_state=42 \\ 
		\hline
	\end{tabular}
	\caption{Parameter Settings for Each Model}
		\label{tab:parameters}
\end{table}

The specific features selected for this experiment included the number of references, citations over a three-year period, and citation structural diversity values, which were used to predict citations for the target literature over 1 year, 5 years, and 10 years, respectively. To compare the impact of citation structural diversity, a baseline model was created to make predictions using only data that did not include citation structural diversity features. The experimental results obtained using the set parameters are shown in Table \ref{tab:model_prediction_results}:

\begin{table}[htbp]
	\centering
	\resizebox{\textwidth}{!}{ 
		\begin{tabular}{|p{1cm}|p{1cm}|p{1cm}|p{1cm}|p{1.5cm}|p{1.5cm}|p{1.5cm}|p{1.5cm}|p{1.5cm}|p{1.5cm}|p{1.5cm}|p{1.5cm}|p{1.5cm}|p{1.5cm}|p{1.5cm}|p{1.5cm}|} 
			\hline
			\multirow{4}{*}{\textbf{Model}} & \multirow{4}{*}{\textbf{T}} &  \multicolumn{6}{l|}{\textbf{Traditional citation network model}} & \multicolumn{8}{l|}{\textbf{Enhanced citation network model}} \\
			\hline
			
			& & \multicolumn{2}{l|}{\textbf{\parbox{2cm}{\raggedright \setlength{\spaceskip}{0.5em}Baseline model}}} 
			& \multicolumn{2}{l|}{\textbf{\parbox{3cm}{\raggedright \setlength{\spaceskip}{0.5em}Citation network model}}} 
			& \multicolumn{2}{l|}{\textbf{\parbox{4cm}{\raggedright \setlength{\spaceskip}{0.5em}Combined citation network model}}} 
			& \multicolumn{2}{l|}{\textbf{\parbox{4cm}{\raggedright \setlength{\spaceskip}{0.5em}Semantic-enhanced citation network model}}} 
			& \multicolumn{2}{l|}{\textbf{\parbox{4cm}{\raggedright \setlength{\spaceskip}{0.5em}Combined-enhanced citation network model}}} 
			& \multicolumn{2}{l|}{\textbf{\parbox{4cm}{\raggedright \setlength{\spaceskip}{0.5em}Semantic-combined-enhanced citation network model}}} 
			& \multicolumn{2}{l|}{\textbf{\parbox{4cm}{\raggedright \setlength{\spaceskip}{0.5em}Combined-semantic-enhanced citation network model}}} \\
			\hline

			& & \textbf{\( R^{2} \)} & \textbf{MSE} & \textbf{\( R^{2} \)} & \textbf{MSE} & \textbf{\( R^{2} \)} & \textbf{MSE} & \textbf{\( R^{2} \)} & \textbf{MSE} & \textbf{\( R^{2} \)} & \textbf{MSE} & \textbf{\( R^{2} \)} & \textbf{MSE} & \textbf{\( R^{2} \)} & \textbf{MSE} \\

			\hline
			\multirow{3}{*}{\textbf{LR}} & 1 & 0.786 & 4.079 & 0.788 & 4.041 & \textbf{\textit{0.788}} & \textbf{\textit{4.028}} & 0.606 & 5.785 & 0.788 & 4.041 & 0.787 & 4.063 & 0.787 & 4.061 \\
			\hline
			
			& 5 & 0.937 & 26.303 & 0.937 & 26.393 & \textbf{\textit{0.938}} & \textbf{\textit{26.149}} & 0.903 & 65.255 & 0.937 & 26.232 & 0.937 & 26.246 & 0.937 & 26.286 \\
			\hline
			
			& 10 & 0.713 & 356.509 & 0.713 & 357.142 & \textbf{\textit{0.715}} & \textbf{\textit{354.352}} & 0.616 & 1474.288 & 0.714 & 355.385 & 0.715 & 354.554 & 0.715 & 354.568 \\
			\hline
			
			\multirow{3}{*}{\textbf{KNN}} & 1 & 0.732 & 5.098 & 0.731 & 5.116 & 0.725 & 5.237 & \textbf{\textit{0.775}} & \textbf{\textit{3.296}} & 0.730 & 5.142 & 0.731 & 5.127 & 0.724 & 5.253 \\
			\hline
			
			& 5 & 0.943 & 23.761 & 0.938 & 26.143 & 0.921 & 33.195 & 0.850 & 101.466 & \textbf{\textit{0.944}} & \textbf{\textit{23.531}} & 0.941 & 24.583 & 0.933 & 28.282 \\
			\hline
			
			& 10 & 0.735 & 329.86 & 0.719 & 349.729 & 0.664 & 417.776 & 0.563 & 1677.667 & \textbf{\textit{0.740}} & \textbf{\textit{322.761}} & 0.732 & 332.463 & 0.727 & 339.446 \\
			\hline
			
			\multirow{3}{*}{\textbf{SVR}} & 1 & 0.418 & 11.084 & 0.471 & 10.078 & 0.553 & 8.507 & 0.425 & 8.435 & 0.521 & 9.119 & \textbf{\textit{0.598}} & \textbf{\textit{7.655}} & 0.584 & 7.928 \\
			\hline
			
			& 5 & 0.580 & 175.92 & 0.603 & 166.577 & 0.687 & 131.293 & 0.717 & 191.217 & 0.672 & 137.567 & \textbf{\textit{0.733}} & \textbf{\textit{111.98}} & 0.721 & 116.907 \\
			\hline
			
			& 10 & 0.437 & 699.959 & 0.433 & 703.966 & 0.494 & 628.373 & 0.588 & 1579.878 & 0.476 & 650.85 & \textbf{\textit{0.535}} & \textbf{\textit{578.107}} & 0.508 & 610.78 \\
			\hline
			
			\multirow{3}{*}{\textbf{CART}}  & 1 & 0.683 & 6.042 & 0.747 & 4.815 & \textbf{\textit{0.763}} & \textbf{\textit{4.518}} & 0.709 & 4.272 & 0.742 & 4.908 & 0.698 & 5.741 & 0.711 & 5.506 \\
			\hline
			
			& 5 & 0.874 & 52.918 & 0.859 & 59.154 & 0.812 & 78.914 & 0.872 & 86.649 & 0.909 & 38.321 & \textbf{\textit{0.917}} & \textbf{\textit{34.805}} & 0.877 & 51.708 \\
			\hline
			
			& 10 & \textbf{\textit{0.679}} & \textbf{\textit{398.58}}1 & 0.563 & 542.74 & 0.570 & 534.386 & 0.627 & 1430.784 & 0.548 & 561.141 & 0.465 & 664.199 & 0.675 & 403.394 \\
			\hline
			
			\multirow{3}{*}{\textbf{MLP}} & 1 & 0.681 & 6.080 & 0.746 & 4.839 & 0.759 & 4.592 & \textbf{\textit{0.796}} & \textbf{\textit{2.997}} & 0.762 & 4.539 & 0.758 & 4.598 & 0.759 & 4.586 \\
			\hline
			
			& 5 & \textbf{\textit{0.925}} & \textbf{\textit{31.387}} & 0.871 & 54.035 & 0.915 & 35.515 & 0.836 & 110.629 & 0.906 & 39.423 & 0.922 & 32.572 & 0.897 & 43.130 \\
			\hline
			
			& 10 & 0.653 & 430.745c & 0.744 & 318.406 & \textbf{\textit{0.760}} & \textbf{\textit{298.164}} & 0.532 & 1796.321 & 0.699 & 374.032 & 0.730 & 335.326& 0.648 & 437.92 \\
			\hline
		\end{tabular}
	}
	\caption{Model Performance Evaluation with Different Citation Networks}
	\label{tab:model_prediction_results}
	
\end{table}

In selecting the optimal model combinations, although some of the combinations are not optimal in terms of a single metric, by combining the two metrics of coefficient of determination \( R^{2} \) and mean squared error (MSE), the final models selected are optimal in terms of overall performance. We highlight the results of the optimal model in bold italicized form. The experimental results show that the inclusion of citation structural diversity features generally improves model prediction, especially for 5- and 10-year citation predictions, where the improvement in model performance is more significant. The enhanced citation network model is overall better than the traditional model in terms of prediction accuracy. By incorporating citation structural diversity features, the \( R^{2} \) values of the enhanced models generally increased, while the MSE values decreased significantly. Additionally, the prediction effect varies across different time windows. For short-term predictions within 1 year, the model generally performs poorly; however, as the time span increases, the \( R^{2} \) values improve significantly. This suggests that longer time spans provide more valid information for the model, significantly improving prediction performance. This further indicates that in citation prediction, data from longer time windows help reveal the long-term academic impact of literature and improve the reliability and accuracy of predictions.

\subsection*{Comparison Experiments of Different Semantic Models}
In our experiments, we compared the performance of three embedding models (Specter \cite{ref30}, BERT \cite{ref28}, and SciBERT \cite{ref29}) in calculating the Pearson correlation coefficients between citation structural diversity and citations under the semantic-combined-enhanced citation network model. The same threshold settings were applied to all three models during the computation to ensure fairness in the comparison. The aim of this comparison is to assess the ability of different embedding models to capture the correlation between citation structural diversity and citation volume.

In analyzing the Pearson correlation coefficients between citation structural diversity and citations, we used two citation calculations: the median of citations and the IQRMean of citations. The specific results for each group of experiments are shown in Table \ref{tab:comparison_results}, where the best-performing results for each group are highlighted in bold.

\begin{table}[ht]
	\centering
	
	\begin{tabular}{|l|l|l|l|l|l|l|}  
	\hline
		& \multicolumn{2}{l|}{\footnotesize Specter} & \multicolumn{2}{l|}{\footnotesize BERT} & \multicolumn{2}{l|}{\footnotesize SciBERT} \\
		\cline{2-7}  
		& \footnotesize {median} & \footnotesize {IQRMean} & \footnotesize {median} & \footnotesize {IQRMean} & \footnotesize {median} & \footnotesize {IQRMean}\\
		\hline
		
		\footnotesize {Group 1} & \footnotesize 0.783 & \footnotesize \textbf{0.786} & \footnotesize -0,831 & \footnotesize -0.657 & \footnotesize -0.831 & \footnotesize -0.593 \\
		\hline
		\footnotesize {Group 2} & \footnotesize 0.828 & \footnotesize \textbf{0.922} & \footnotesize 0.917 & \footnotesize 0.909 & \footnotesize 0.621 & \footnotesize 0.639 \\
		\hline
		\footnotesize {Group 3} & \footnotesize 0.637 & \footnotesize 0.747 & \footnotesize 0.705 & \footnotesize \textbf{0.808} & \footnotesize 0.736 & \footnotesize \textbf{0.831} \\
		\hline
		\footnotesize {Group 4} & \footnotesize 0.421 & \footnotesize 0.452 & \footnotesize 0.609 & \footnotesize \textbf{0.616} & \footnotesize 0.403 & \footnotesize 0.443 \\
		\hline
		\footnotesize {Group 5} & \footnotesize 0.798 & \footnotesize \textbf{0.869} & \footnotesize 0.500 & \footnotesize 0.537 & \footnotesize 0.534 & \footnotesize 0.561 \\
		\hline
		\footnotesize {Group 6} & \footnotesize 0.649 & \footnotesize 0.642 & \footnotesize \textbf{0.882} & \footnotesize 0.875 & \footnotesize 0.682 & \footnotesize 0.677 \\
		\hline
		\footnotesize {Group 7} & \footnotesize 0.649 & \footnotesize \textbf{0.677} & \footnotesize 0.437 & \footnotesize 0.481 & \footnotesize 0.558 & \footnotesize 0.591 \\
		\hline
		\footnotesize {Mean} & \footnotesize 0.681 & \footnotesize \textbf{0.728} & \footnotesize 0.460 & \footnotesize 0.510 & \footnotesize 0.386 & \footnotesize 0.450 \\
		\hline
		\footnotesize {Variance} & \footnotesize \textbf{0.017} & \footnotesize 0.021 & \footnotesize 0.305 & \footnotesize 0.251 & \footnotesize 0.257 & \footnotesize 0.193 \\
		\hline
	\end{tabular}

	\caption{Results of Model Comparison Experiments}
	\label{tab:comparison_results}
\end{table}

It is worth noting that some experimental groups showed negative results, suggesting a negative correlation between citation structural diversity and citations in certain datasets. A negative correlation does not necessarily imply that the model is invalid but reflects issues with the data characteristics themselves or the adaptability of the model. For instance, certain fields or types of literature may have low citations but high citation structural diversity, or vice versa, which could lead to negative correlation results.

As seen from the table, the Specter model outperforms the BERT and SciBERT models in terms of both the mean and variance of the Pearson correlation coefficients for both citation volume calculations. Specifically, the Specter model has a higher mean correlation coefficient and lower variance, indicating greater stability and reliability in capturing the correlation between citation structural diversity and citation volume. In contrast, the BERT and SciBERT models show lower mean correlation coefficients and higher variance, with the BERT model particularly exhibiting a significantly higher variance in the correlation coefficients compared to the Specter model.

\subsection*{The Impact of citation structural diversity on the Measurement of Interdisciplinarity in Literature}

Effective literature evaluation metrics, especially those related to highly innovative and interdisciplinary papers, can provide valuable insights for researchers, broadening their research perspectives and enhancing their innovation capabilities \cite{ref41}. To explore whether the citation structural diversity of literature is related to interdisciplinarity, we designed and conducted correlation experiments. However, preliminary analysis revealed that the original dataset lacked detailed topic classification information, which limited in-depth exploration of interdisciplinarity. Therefore, to obtain more comprehensive data, we decided to use the PubMed dataset. Unlike the previous dataset, PubMed not only includes rich bibliographic information but also provides detailed topic classifications, greatly facilitating interdisciplinary analysis.

First, we conducted thorough data cleaning and preprocessing on the PubMed dataset to ensure the timeliness and relevance of the study. We selected literature published in 2020 to avoid the issues of insufficient information in older publications and the potential data shortage caused by a low number of recent publications. Afterward, we classified the literature based on the ranking of publication. Upon completing the classification, we randomly selected four representative groups of data from different publication types to reduce sample bias and ensure the generalizability of the results. Each group contained a different number of literatures to reflect the relationship between citation structural diversity and interdisciplinarity in the subsequent analysis. The specific data are shown in Table \ref{tab:literature_Information} below:

\begin{table}[ht]
	\centering
	
	\renewcommand{\arraystretch}{1.5}  
	\begin{tabular}{|>{\raggedright\arraybackslash}m{4cm}| >{\raggedright\arraybackslash}m{6cm}|  >{\raggedright\arraybackslash}m{2cm}|} 
		\hline
		\textbf{Name of Publication} & \textbf{Type of Publication}  & \textbf{Number of literatures} \\ 
		\hline
		Group 1 & International Journal of Biomedical Computing and Medical Informatics Decision Making  & 292 \\ 
		\hline
		Group 2 & BMJ Case Reports & 408 \\ 
		\hline
		Group 3 & Elife & 763 \\ 
		\hline
		Group 4 & mBio & 548 \\ 
		\hline
	\end{tabular}
	\caption{Specific Information on Seven Groups of Literature from the Same Year and Journal}
	\label{tab:literature_Information}
\end{table}

After completing the data preparation, we applied the same analysis model used in previous experiments to calculate the citation structural diversity of the literature in each group. Using the pre-established model, we calculated the citation structural diversity value for each paper based on citation relationships and semantic relationships. citation structural diversity reflects the multifaceted connectivity of literatures in the citation network. During the calculation of citation structural diversity, we used different thresholds compared to previous experiments: when adding semantic edges, the lower quartile value of the edge between the target paper and the reference paper was used as the threshold \( \theta_1 \); while when adding combined edges, the IQRMean of the edge between the target paper and the reference paper was used as the threshold \( \theta_2 \).

At the same time, we extracted the topic count information for each paper from the dataset as an important indicator of interdisciplinarity, which intuitively reflects the breadth of the academic fields covered by the paper. To quantify the relationship between citation structural diversity and the number of topics, we calculated the Pearson correlation coefficient (\( r \)) between the two. Using this method, we obtained the Pearson correlation coefficients between citation structural diversity and the number of topics for different models. The results of the correlation analysis are presented in Table \ref{tab:correlation_results}:

\begin{table}[ht]
	\centering
	\footnotesize
	\resizebox{\textwidth}{!}{
		\begin{tabular}{|>{\raggedright\arraybackslash}p{1.5cm}| 
				>{\raggedright\arraybackslash}p{2.0cm}| 
				>{\raggedright\arraybackslash}p{3.0cm}| 
				>{\raggedright\arraybackslash}p{3.5cm}| 
				>{\raggedright\arraybackslash}p{3.5cm}| 
				>{\raggedright\arraybackslash}p{4cm}|
				>{\raggedright\arraybackslash}p{4cm}|}
			\hline
			\textbf{Group} & 
			\textbf{Citation Network Model} & 
			\textbf{Combined Citation Network Model} & 
			\textbf{Semantic-Enhanced Citation Network Model} & 
			\textbf{Combined-Enhanced Citation Network Model} & 
			\textbf{Semantic-Combined-Enhanced Citation Network Model} & 
			\textbf{Combined-Semantic-Enhanced Citation Network Model} \\
			\hline
			Group 1 & 0.365 & 0.055 & 0.368 & \textbf{\textit{0.430}} & 0.406 & 0.071 \\
			\hline
			Group 2 & 0.427 & 0.124 & \textbf{\textit{0.492}} & 0.476 & 0.476 & 0.119 \\
			\hline
			Group 3 & 0.608 & \textbf{\textit{0.649}} & 0.622 & 0.609 & 0.622 & 0.464 \\
			\hline
			Group 4 & 0.387 & 0.093 & \textbf{\textit{0.499}} & 0.454 & 0.496 & 0.069 \\
			\hline
			\textbf{Mean} & 0.447 & 0.230 & 0.495 & 0.492 & \textbf{\textit{0.500}} & 0.181 \\
			\hline
			\textbf{Variance} & 0.009 & 0.059 & 0.008 & \textbf{\textit{0.005}} & 0.006 & 0.027 \\
			\hline
		\end{tabular}
	}
	\caption{Pearson Correlation Coefficients Between citation structural diversity and the Number of Topics} 
	\label{tab:correlation_results}
\end{table}

Based on the analysis of Pearson correlation coefficients, we found a significant correlation between structural diversity and interdisciplinarity. Specifically, changes in structural diversity across different literature groups and models were found to correlate with the breadth of academic fields covered by the literature. In particular, papers with higher structural diversity values tended to cover a broader range of academic fields, indicating stronger interdisciplinary characteristics. This finding validates the effectiveness of structural diversity as a key indicator for measuring the interdisciplinarity of literature. Additionally, the introduction of enhanced models (such as the Semantic-Combined Enhanced Citation Network Model) significantly improved the accuracy of the correlation analysis, further demonstrating the advantages of the proposed model in assessing interdisciplinarity.
\section*{Conclusion}

This study combines citation structural features and semantic information through the innovative design of a citation structural diversity model, proposing a literature evaluation method. The results show that citation structural diversity is valuable for measuring the academic impact and identifying interdisciplinarity, which provides new technical support and theoretical references for scientific evaluation.

Despite these important findings, some aspects of the study warrant further exploration. For instance,  the model’s generalizability in different academic contexts needs further validation to ensure its broad utility. Moreover, the dynamic nature of citation structural diversity at the citing literature end, as well as its real-time impact on scholarly dissemination, has not been thoroughly examined. Future research should address these areas to deepen our understanding of how citation structural diversity evolves over time and its direct influence on academic communication.

\bibliography{sn-bibliography}  

\section*{Author contributions} Mingyue Kong contributed to the conceptualization, methodology, data curation, software development, and the writing of the original draft, as well as the review and editing of the manuscript. \\ Yinglong Zhang was responsible for the conceptualization, methodology, data curation, supervision, and the writing of the original draft, in addition to the review and editing of the manuscript. \\ Likun Sheng provided supervision and contributed to the review and editing of the manuscript. \\ Kaifeng Hong also provided supervision and contributed to the review and editing of the manuscript.

\section*{Competing interest}
No conflict of interest exists in the submission of this manuscript, and the manuscript is approved by all authors for publication.

\section*{Funding}
This work was supported by the Natural Science Foundation of Fujian Province of China (No. 2023J01922); the Advanced Training Program of Minnan Normal University (No. MSGJB2023015); the Headmaster Fund of Minnan Normal University (No. KJ19009); Zhangzhou City’s Project for Introducing High-level Talents; and the National Natural Science Foundation of China (No. 61762036).
 62163016.

\section*{Additional Information}
\noindent \textbf{Correspondence} and requests for materials should be addressed to Y.Z \\
\noindent \textbf{Reprints and permissions information} is available at \url{www.nature.com/reprints}. \\
\noindent \textbf{Publisher’s note} Springer Nature remains neutral with regard to jurisdictional claims in published maps and institutional affiliations. \\
\noindent \textbf{Open Access} This article is licensed under a Creative Commons Attribution-NonCommercial-NoDerivatives 4.0 International License, which permits any non-commercial use, sharing, distribution and reproduction in any medium or format, as long as you give appropriate credit to the original author(s) and the source, provide a link to the Creative Commons licence, and indicate if you modified the licensed material. You do not have permission under this licence to share adapted material derived from this article or parts of it. The images or other third party material in this article are included in the article’s Creative Commons licence, unless indicated otherwise in a credit line to the material. If material is not included in the article’s Creative Commons licence and your intended use is not permitted by statutory regulation or exceeds the permitted use, you will need to obtain permission directly from the copyright holder. To view a copy of this licence, visit \url{http://creativecommons.org/licenses/by-nc-nd/4.0/}.

\end{document}